\documentclass[reprint,amsmath,amssymb,aps,twocolumn,superscriptaddress]{revtex4-2}
\usepackage{graphicx}
\usepackage{dcolumn}
\usepackage{bm}
\usepackage{float}
\usepackage{amssymb}
\usepackage{color}
\usepackage{multirow}
\usepackage{stmaryrd}
\usepackage{natbib}
\usepackage[colorlinks,linkcolor=blue,urlcolor=blue,citecolor=blue]{hyperref}

\begin{document}


\title{Mn$_4$Al$_{11}$: A Half-Semimetal Candidate with Anomalous Electronic Behaviors}

\author{Shanshan Han}
\thanks{These authors contributed equally to this work.}
\affiliation{Beijing Academy of Quantum Information Sciences, Beijing 100193, China}
\affiliation{School of Materials Science and Engineering, Nankai University, Tianjin 300350, China}

\author{Rongsheng Li}
\thanks{These authors contributed equally to this work.}
\affiliation{International Center for Quantum Materials, School of Physics, Peking University, Beijing 100871, China}
\author{Xingxing Jiang}
\affiliation{Department of Applied Physics, School of Physics and Electronics, Hunan University, Changsha 410082, China}
\author{Ming Lu}
\affiliation{Beijing Academy of Quantum Information Sciences, Beijing 100193, China}

\author{Shuxiang Xu}
\affiliation{International Center for Quantum Materials, School of Physics, Peking University, Beijing 100871, China}

\author{Jiang Zeng}
\affiliation{Department of Applied Physics, School of Physics and Electronics, Hunan University, Changsha 410082, China}
\author{Dong Wu}
\email{wudong@baqis.ac.cn}
\affiliation{Beijing Academy of Quantum Information Sciences, Beijing 100193, China}

\author{Nanlin Wang}
\email{nlwang@pku.edu.cn}
\affiliation{Beijing Academy of Quantum Information Sciences, Beijing 100193, China}
\affiliation{International Center for Quantum Materials, School of Physics, Peking University, Beijing 100871, China}
\affiliation{Collaborative Innovation Center of Quantum Matter, Beijing 100871, China}

\date{\today}

\begin{abstract}
Half-semimetals, characterized by their spin-polarized electronic states, hold significant promise for spintronic applications but remain scarce due to stringent electronic and magnetic criteria. Through a combination of transport measurements and optical spectroscopy, we investigated the intermetallic compound Mn$_4$Al$_{11}$, which features an exceptionally low carrier concentration and undergoes a magnetic phase transition near 68 K. Transport measurements reveal anomalies that deviate from typical metallic behavior at low temperatures. Optical spectroscopy indicates a small, nearly frequency-independent optical conductivity in the far-infrared region, with spectral weight decreasing as the temperature drops from 300 K to 50 K. These behaviors suggest a temperature-dependent carrier density and significant scattering of charge carriers. Combining experimental findings with calculated electronic band structures, we propose that Mn$_4$Al$_{11}$ is a novel half-semimetal candidate exhibiting a ferrimagnetic ground state.
\end{abstract}

\pacs{Valid PACS appear here}
\maketitle


\section{\label{sec:level1}INTRODUCTION}

Intermetallic compounds constitute a distinct class of metallic alloys characterized by ordered crystal structures with well-defined stoichiometric ratios between the constituent metals. These materials have garnered extensive attention in condensed matter physics and materials science due to their unique electronic configurations and diverse functional properties, including superconductivity~\cite{cava1994superconductivity2,he2001superconductivity}, topological Kagome magnets~\cite{ye2018massive, teng2022discovery}, thermoelectricity~\cite{zlatic2005theory}, catalytic reactions~\cite{furukawa2017intermetallic}, and hydrogen storage capability~\cite{kamakoti2003comparison}. Among them, transition-metal-based intermetallics have attracted particular attention due to their complex magnetic phase diagrams and correlated electron behaviors. In particular, these materials may host exotic semimetallic or semiconducting states~\cite{buschow1977intermetallic,  petrova2010elastic, martin2023optical}. Few known examples, such as FeSi, highlight both the challenges and opportunities in this research area~\cite{schlesinger1993unconventional, paschen1997low,fath1998tunneling}. FeSi demonstrates a narrow-gap semiconducting ground state below 100 K, exhibiting electronic properties that deviate significantly from those predicted by a simple band model~\cite{paschen1997low,fath1998tunneling}. Similar anomalies emerge in FeSb$_2$ and FeGa$_3$, which feature exotic electronic and spin behavior and are also considered potential heavy fermion Kondo insulators~\cite{Bentien_2007, Bittar_2010}.

\begin{figure}[htbp]
\includegraphics[clip, width=0.5\textwidth]{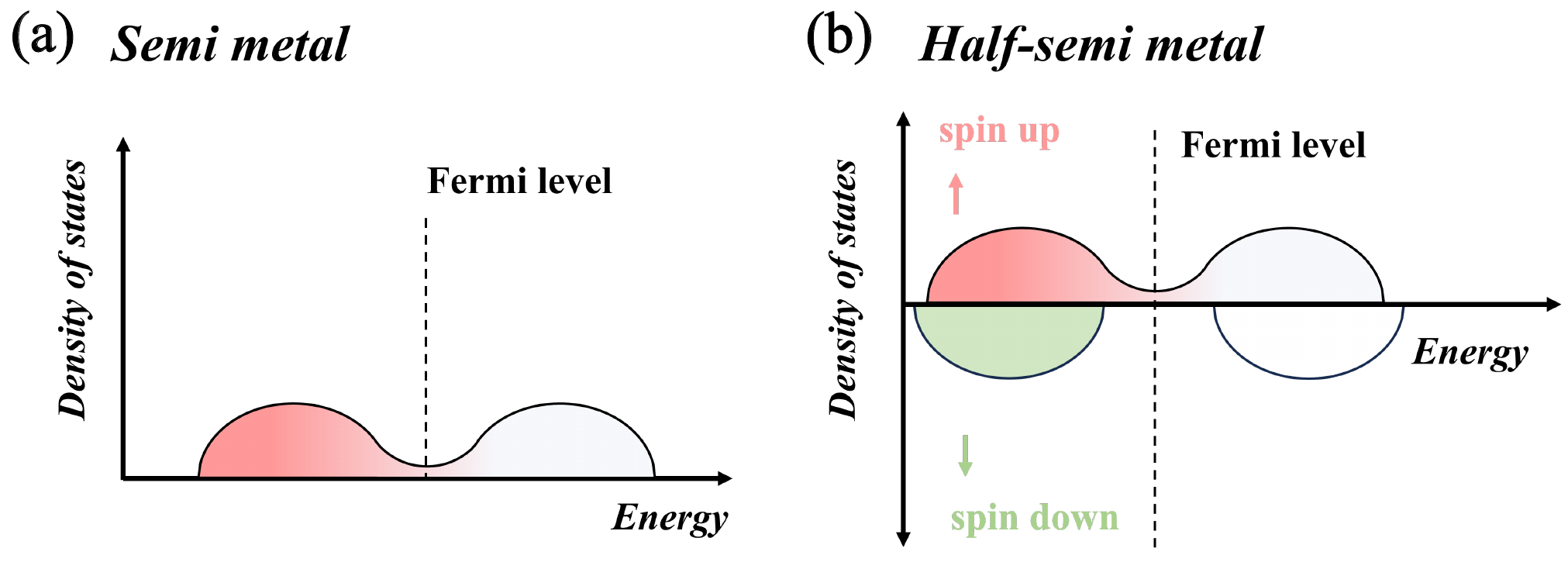}\\[1pt] 
\caption{{The electronic structure schematics of semimetal (a) and half-semimetal (b).}\label{fig:1}}
\end{figure}

In addition to those exotic semiconducting and semimetallic behaviors, certain intermetallic compounds exhibit even more intriguing electronic characteristics, such as half-semimetals. This distinctive class of materials is characterized by spin-selective band gaps, enabling fully spin-polarized conduction channels. As illustrated in Fig.~\ref{fig:1}(a), traditional semimetals exhibit a slight overlap between conduction and valence bands at the Fermi level. In contrast, half-semimetals feature a gapless conduction band for one spin orientation while exhibiting a band gap for the opposite spin orientation~\cite{Coey2004Magnetic, RevModPhys.80.315} (Fig.~\ref{fig:1}(b)). Such spin-filtering capability positions half-semimetals as a novel playground for spintronics~\cite{Bowen2003Nearly, PhysRevB.62.R4790}.

\begin{figure*}[htbp]
\includegraphics[clip, width=1\textwidth]{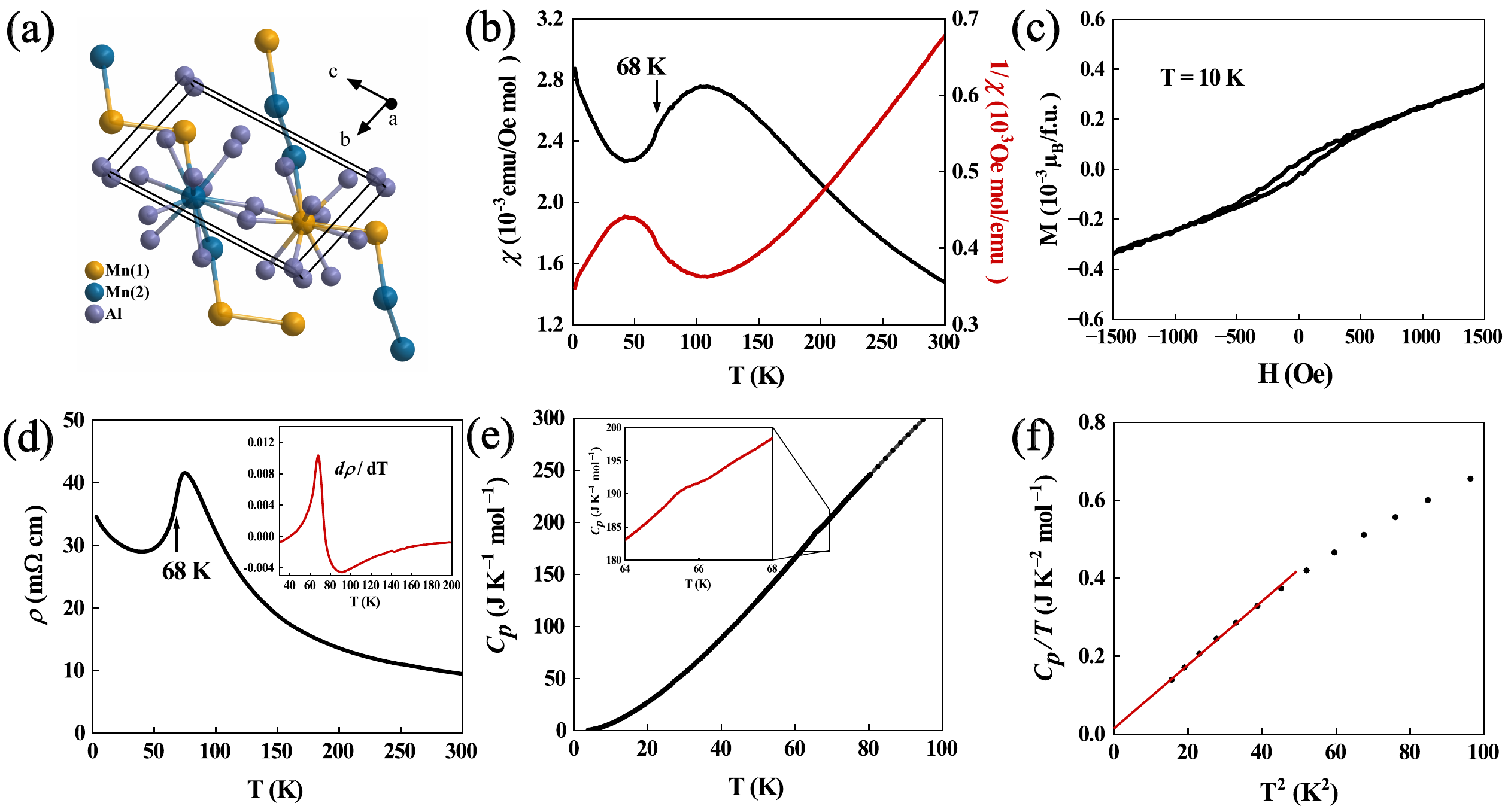}\\[1pt] 
\caption{
Crystal structure, magnetic susceptibility, resistivity, and specific heat of Mn$_4$Al$_{11}$. 
(a) Crystal structure of Mn$_4$Al$_{11}$ within a unit cell (gray lines). 
(b) Temperature-dependent magnetic susceptibility and inverse susceptibility measured under zero-field cooling at $B = 0.1$~T with $B \parallel c$. 
(c) Magnetic hysteresis loop measured at 10~K with the magnetic field applied along the $c$-axis. 
(d) Temperature dependence of the dc resistivity. The inset shows the temperature derivatives of the resistivity. The black arrows mark the phase transition at 68~K.
(e) Specific heat capacity showing a weak anomaly at 68~K. 
(f) Low-temperature $C_p/T$ fitted with the Sommerfeld model (red curve). 
}
\label{fig:2}
\end{figure*}

Here, we report on a Mn-based transition metal intermetallic compound Mn$_4$Al$_{11}$. According to the literature, Mn$_4$Al$_{11}$ exhibits intriguing physical properties, including a negative temperature dependence of resistivity and possible one-dimensional magnetism~\cite{Kontio1980new, Dunlop1976one, PhysRevB.109.064207}. Although the structure, magnetic susceptibility, resistivity, and band structure of Mn$_4$Al$_{11}$ have been reported, the mechanism of these exotic transport behaviors remains underexplored. We systematically characterized and investigated the underlying physics of  Mn$_4$Al$_{11}$ by combining transport measurements, optical spectroscopy, and first-principles calculations. The temperature-dependent resistivity exhibits a semiconductor-like behavior above 100 K, with a distinct anomaly emerging at 68 K that correlates with magnetic susceptibility data. Below this transition temperature, both the resistivity and the magnetic moment exhibit significant quenching. The far-infrared real part of conductivity reveals a low spectral weight at room temperature, decreasing with cooling, indicating an extremely low carrier concentration down to low temperature. Crucially, our first-principles calculations identify Mn$_4$Al$_{11}$ as a promising candidate for a new type of half-semimetal.

\section{\label{sec:level2}RESULTS AND DISCUSSION}

\begin{figure*}[htbp]
\includegraphics[clip, width=0.8\textwidth]{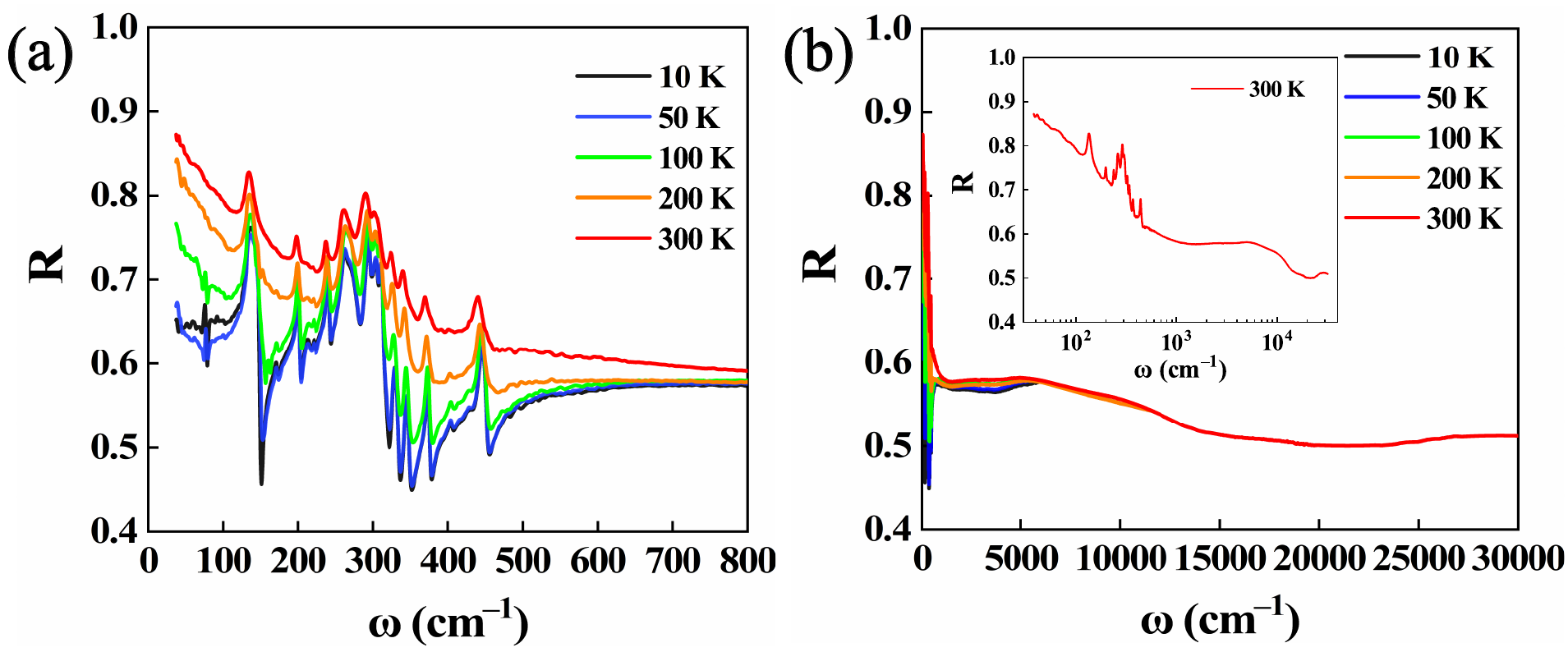}\\[1pt] 
\caption{Temperature-dependent optical reflectivity R($\omega$) in the far-infrared region (a) and in an expanded region up to 10000 cm$^{-1}$(b). 
\label{fig:3}}
\end{figure*}
High-quality single crystals of Mn$_4$Al$_{11}$ were grown using the Sn flux method. High-purity Mn (99.99\%), Al (99.99\%), and Sn (99.99\%) were sealed in an evacuated quartz tube with a molar ratio of Mn:Al:Sn = 3:10:20. The tube was heated from room temperature to 1000 $^{\circ}$C over 15 h in a muffle furnace, held at this temperature for 50 h, and then cooled down to 620 $^{\circ}$C at 1.5 $^{\circ}$C/h. After removing the flux by centrifugation, shiny, plate-like Mn$_4$Al$_{11}$ crystals were obtained.

The crystal structure of Mn$_4$Al$_{11}$ was determined by X-ray diffraction data collected using a Rigaku XtaLAB PRO 007HF diffractometer equipped with graphite-monochromated Mo K$_\alpha$ radiation ($\lambda$ = 0.71073 $\mathrm{\AA}$) at 92 K. The structure was solved by direct methods and refined on $F^2$ using full-matrix least-squares via the SHELXTL2014 crystallographic software package. The corresponding crystallographic information file (CIF) has been deposited with the Cambridge Crystallographic Data Centre under accession number CCDC 2414038.

Mn$_4$Al$_{11}$ adopts a centrosymmetric triclinic structure with the space group $P\overline{1}$. There are two Mn and six Al atoms in the crystallographically independent unit in the structure of Mn$_4$Al$_{11}$, as illustrated in Fig.~\ref{fig:2}(a). Each Mn atom is coordinated by two Mn and ten Al atoms forming a triclinic structure. Two Mn(1) and two Mn(2) atoms are arranged in an overlapping manner, forming a 1D zigzag chain along the $c$-axis. The distances between adjacent Mn(1)-Mn(1), Mn(1)-Mn(2), and Mn(2)-Mn(2) atoms are 3.207 $\mathrm{\AA}$, 3.193 $\mathrm{\AA}$, and 3.079 $\mathrm{\AA}$, respectively. The distance between the two chains is relatively large, indicating a weak interaction between them.

\begin{figure*}[htbp]
\includegraphics[clip, width=1\textwidth]{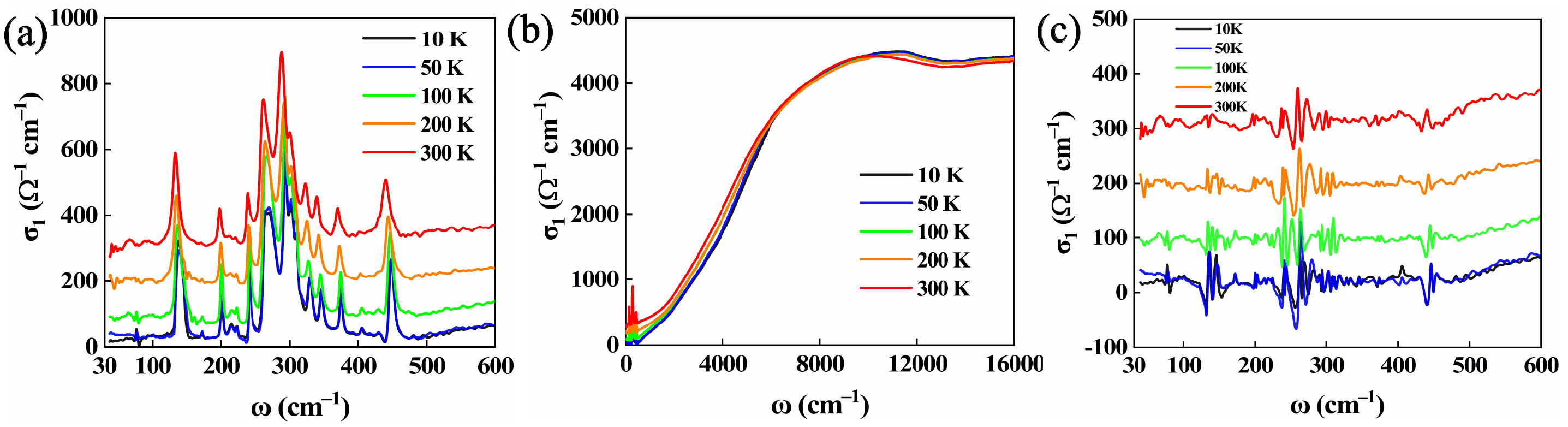}\\[1pt] 
\caption{The experimental optical conductivity $\sigma_1$($\omega$) at different temperatures in the far-infrared region (a) and in an expanded region up to 10000 cm$^{-1}$(b). 
(c) The experimental optical conductivity $\sigma_1$($\omega$) at different temperatures with phonon structures being subtracted in the far-infrared region. Nevertheless, because of the abundance of phonons, it is challenging to remove them completely.
\label{fig:4}}
\end{figure*}

We performed magnetic susceptibility, resistivity, and heat capacity measurements on single crystals of Mn$_4$Al$_{11}$. 
The electrical resistivity was measured using a Physical Property Measurement System (PPMS) with a standard four-probe configuration. Gold wires and silver epoxy were used to form electrical contacts. The sweep rate was set to 2 K/min unless otherwise specified. Magnetic measurements (DC susceptibility) were performed using a Quantum Design Magnetic Property Measurement System (MPMS) with the magnetic field applied along the $c$-axis. For a typical temperature sweep experiment, the sample was cooled to 2 K under zero-field cooled (ZFC) conditions, and data were collected by warming from 2 to 300 K under a field of 1000 Oe. Isothermal magnetization as a function of magnetic field ($M$–$H$) was measured at 10 K with the magnetic field also applied along the $c$-axis, sweeping the field from –6 T to +6 T. Specific heat $C_p$(T) measurements on single crystals were also performed using the Quantum Design PPMS via the relaxation method. Apiezon N grease was used to ensure good thermal and mechanical contact between the sample and the heat capacity stage.

By analyzing the temperature derivatives of the transport data, we can identify a phase transition temperature \emph{T$^{*}$} = 68 K. Specifically, the temperature-dependent magnetization for Mn$_4$Al$_{11}$ is found to be positive and increases rapidly with decreasing temperature above ~100 K, then decreases to 50 K [Fig.~\ref{fig:2}(b)]. This decrease maybe be attributed to the enhanced antiferromagnetic-type correlations within the Mn(1)–Mn(2) zigzag chains. Below \emph{T$^{*}$}, the magnetic moment gradually decreases, signaling the emergence of antiferromagnetic ordering. 
To further quantify the magnetic behavior, we fit the inverse susceptibility above 200 K using the Curie-Weiss model [Fig.~\ref{fig:2}(b)]. The linear fit [$\chi = \frac{C}{T - \theta_w}$] yields a Weiss temperature $\theta_w = -45$ K, a Curie constant $C = 0.53$ emu·K·Oe$^{-1}$·mol$^{-1}$, and an effective magnetic moment $\mu_{\mathrm{eff}} \approx 1.96\,\mu_B$. While the Curie–Weiss analysis offers valuable information on the magnetic interactions, it cannot determine the exact spin configuration, highlighting the need for further studies such as neutron diffraction or spin-resolved first-principles calculations. Further measurements reveal the formation of a weak ferrimagnetic ground state in this system [Fig.~\ref{fig:2}(c)], consistent with our density functional theory (DFT) calculations (discussed in detail later) and the reference data from the Materials Project for Mn$_4$Al$_{11}$ (mp-2856, database version v2024.12.18). 



Figure~\ref{fig:2}(d) shows the $ab$-plane dc resistivity curve of Mn$_4$Al$_{11}$ as a function of temperature, similar to the behavior of the magnetic susceptibility. Although the temperature-dependent magnetic susceptibility and resistivity are consistent with previous reports~\cite{Dunlop1976one}, the underlying physics has not been explored. As the temperature decreases from 300 K, $\rho$(T) initially increases, reaching a peak around 75 K, and then decreases to a minimum near 50 K. A sharp peak is observed at approximately 68 K in the resistivity derivative plot shown in the inset of Fig. \ref{fig:2}(d), aligning with the anomaly in the magnetic measurements and indicating a phase transition. Typically, a negative temperature coefficient of resistivity indicates semiconducting behavior, but the complex resistivity behavior does not allow a straightforward classification of the compound as a metal or semiconductor. This is similar to the low carrier density compounds 1$T$-TiSe$_2$ or ZrTe$_5$, where their resistivities also display nonmonotonic temperature dependence, yet optical measurements have confirmed their metallic nature \cite{PhysRevLett.99.027404,PhysRevB.92.075107}. To clarify the issue in Mn$_4$Al$_{11}$, we have conducted optical spectroscopy measurements to determine the low-frequency spectral properties, as we shall present below. The measurements reveal that the compound is a low-carrier-density metal with strong carrier scattering likely originating from local moments formed by localized Mn 3$d$ orbitals. Additionally, the carrier density decreases with temperature, potentially explaining the increase in resistivity as temperature decreases.

The heat capacity measurements shown in Fig.~\ref{fig:2}(e) also indicate a weak entropy anomaly around \emph{T$^*$}. Despite the prominent transition both in the resistivity and magnetization results, the heat capacity anomaly is much weaker. We observe a linear scaling when plotting $C_p/T$ against the square of the temperature $T$ at low temperatures, as shown in Fig.~\ref{fig:2}(f). 
From the Sommerfeld fit [ $\frac{C_p}{T} = \gamma + \frac{12\pi^4}{5} N_0k_B \frac{T^2}{\theta_D^3}$], we estimate the Sommerfeld coefficient $\gamma$= 49 mJ$\cdot$ mol$^{-1}$$\cdot$ K$^{-2}$, which is comparatively larger than typical metals. Furthermore, the Debye temperature is determined to be 112 K. The non-zero electronic specific heat coefficient $\gamma$ suggests that the compound behaves as a metal rather than a semiconductor. We propose that the enhancement of the Sommerfeld coefficient in Mn$_4$Al$_{11}$ is likely due to the correlation effect of Mn 3$d$ electrons, which can be considered the primary factor responsible for the anomalous electron transport observed in the material.

The in-plane reflectivity R($\omega$) was measured using a Bruker 80 V Fourier transform infrared spectrometer over the frequency range from 40 to 32,000 cm$^{-1}$. At low frequencies, an in situ gold overcoating technique was employed to obtain the reflectance data, while aluminum was used as the reference instead of gold above 8000 cm$^{-1}$. 

The temperature-dependent in-plane reflectivity spectra are presented in Fig.~\ref{fig:3}(a) (far-infrared region) and Fig.~\ref{fig:3}(b) (broad energy range).  The reflectivity at 300 K exhibits a clear increase at low frequencies, indicative of metallic behavior. However, its absolute values are significantly lower than those of conventional metals. 
Upon cooling, the low-frequency reflectivity gradually decreases, whereas the response above 5000 cm$^{-1}$ shows minimal temperature dependence. Notably, below 500 cm$^{-1}$, multiple phonon absorption features emerge due to weak electronic screening, and they become more pronounced at lower temperatures.

To further probe the charge dynamics and electronic structure, we derive the real part of the optical conductivity through the Kramers-Kronig transformation, as shown in Fig.~\ref{fig:4}(a) and~\ref{fig:4}(b). The low- and high-frequency extrapolations employed the Hagen-Rubens relation and X-ray atomic scattering functions, respectively. In the low-frequency conductivity spectrum, $\sigma_1$($\omega$) is dominated by strong infrared-active phonon modes, which intensify at lower temperatures.
Below 100 K, additional weak phonon peaks emerge, possibly signaling a phase transition below \emph{T$^*$}. Furthermore, the phonon mode near 300 cm$^{-1}$ displays slight softening at elevated temperatures.

\begin{figure*}[htbp]
\includegraphics[clip, width=0.8\textwidth]{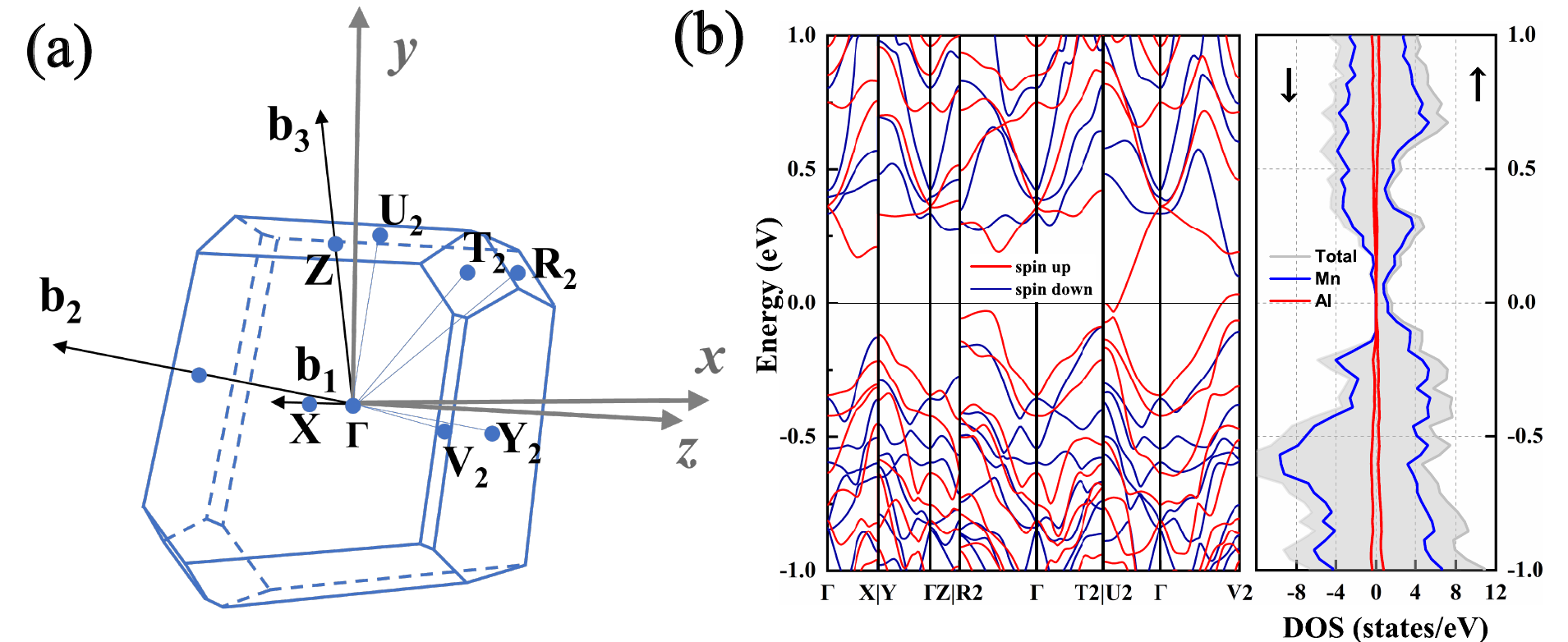}\\[1pt] 
\caption{
(a). Brillouin zone of Mn$_4$Al$_{11}$. (b). Band structures and projected density of states (DOS), where Fermi level is used as a reference point.
\label{fig:5}}
\end{figure*} 
The electronic contribution was isolated by subtracting phonon peaks from the spectrum via least-squares fitting with Lorentzian profiles (Fig.~\ref{fig:4}(c)).  We found that the conductivity still showed finite spectral weight at low frequencies, suggesting that some bands must cross the Fermi level and contribute to the low-frequency conductivity. This confirms that Mn$_4$Al$_{11}$ is a metal rather than an insulator or semiconductor, which is consistent with our analysis of specific heat. Importantly, the low-frequency conductivity does not exhibit a clear Drude-like peak but displays a flat behavior. This may indicate that the mobile carriers experience significant scattering, presumably from the magnetic moments formed by localized Mn 3$d$ electrons. 

When those moments become ordered at \emph{T$^*$}, the scattering becomes smaller, which could explain the sudden reduction of dc resistivity below \emph{T$^*$}. As temperature decreases, the low-frequency flat conductivity further declines, indicating that the mobile carrier density is further reduced. Similar temperature-induced spectral weight decreases were also observed in other low-carrier-density compounds, such as 1T-TiSe$_2$ and ZrTe$_5$ \cite{PhysRevLett.99.027404,PhysRevB.92.075107}. Actually, the separated flat conductivity spectra in $\sigma_1$($\omega$) at reduced temperature are similar to the pseudogap features of YBa$_2$Cu$_3$O$_{6.70}$~\cite{homes1993optical}.
In Al-Mn alloys, $sp$-$d$ hybridization between Al 3$s,p$, and Mn 3$d$ normally depletes the Mn 3$d$ related density of states at E$_F$, creating a pseudogap. Our optical spectra of Mn$_4$Al$_{11}$ are consistent with such the pseudogap scenario~\cite{krajci1997atomic, shukla2008influence}.

We further performed first-principles density functional theory (DFT) calculations to investigate the electronic properties of Mn$_4$Al$_{11}$, using the \textit{Vienna Ab initio Simulation Package} (VASP) \cite{kohn1965self,kresse1994theory}.
We employed projector augmented-wave (PAW) pseudopotentials \cite{mortensen2005real} for core-valence electron interactions and the Perdew-Burke-Ernzerhof (PBE) generalized gradient approximation (GGA) \cite{perdew1996generalized,zhou2017modulation} for exchange-correlation effects. A plane-wave energy cutoff of 500 eV was adopted. The Brillouin zone was sampled using a Monkhorst–Pack $7 \times 7 \times 5$ $k$-point mesh. For structural relaxation, the convergence criterion was set such that the force on each atom was less than 0.02 eV/$\mathrm{\AA}$.

As shown in Fig.~\ref{fig:5}(a), the Brillouin zone of Mn$_4$Al$_{11}$ exhibits a triclinic symmetry with clearly labeled high-symmetry points. The electronic band structure analysis [Fig.~\ref{fig:5}(b)] reveals that the spin-up channel exhibits metallic character, whereas the spin-down channel displays semiconducting nature. The projected density of states (DOS) analysis [Fig.~\ref{fig:5}(b)] confirms that the magnetic Mn atoms are responsible for the metallicity in the spin-up channel through the formation of both electron and hole pockets.  Our calculations support a ferrimagnetic ground state for Mn$_4$Al$_{11}$, consistent with previous proposals in the literature~\cite{ PhysRevB.109.064207}. Moreover, the DOS near the Fermi energy is rather low, consistent with the very low carrier density observed in optical measurements. These results support the classification of Mn$_4$Al$_{11}$ as a half-semimetal. To our knowledge, HgCr$_2$Se$_4$ is currently one of the few materials reported as a half-semimetal \cite{PhysRevLett.107.186806,PhysRevLett.115.087002}, making Mn$_4$Al$_{11}$ an extremely rare compound. Therefore, we rationalize the DFT results in conjunction with our resistivity and optical spectrum experimental observations, which provide fascinating insights into the half-semimetal properties of Mn$_4$Al$_{11}$.

The interplay between the localized magnetic moments and low carrier density can lead to strong electron correlations, potentially giving rise to exotic quantum states. Such systems provide a remarkable platform for investigating heavy fermions, quantum spin liquids, novel topological states, and other quantum phenomena ~\cite{xu2022mechanism,xu2024mechanism,Coey2004Magnetic, RevModPhys.80.315,PhysRevLett.107.186806,PhysRevLett.115.087002}. While our current work focuses on abnormal electronic behavior of Mn$_4$Al$_{11}$, these considerations suggest rich opportunities for future exploration of correlation-driven phenomena in similar material systems.




\section{\label{sec:level3}CONCLUSION}

In summary, we have successfully developed high-quality single crystals of Mn$_4$Al$_{11}$, enabling a comprehensive investigation of its electronic, magnetic, thermodynamic, and optical properties. Our resistivity, magnetic susceptibility, and specific heat measurements reveal that Mn$_4$Al$_{11}$ undergoes a phase transition at 68 K, demonstrating complex magnetic and electronic behavior in the ground state. Notably, the optical conductivity spectra demonstrate a remarkably low carrier concentration, which further decreases at lower temperatures. Moreover, these conducting charge carriers experience rather strong scattering, presumably due to localized moments formed by the localized orbitals of Mn 3$d$ electrons within the zigzag Mn-Mn chains. The DFT calculations further support the characterization of Mn$_4$Al$_{11}$ as a half-semimetal candidate, highlighting its unique electronic structure, with only a limited number of materials exhibiting such behavior. Given the extreme rarity of half-semimetal materials, we believe our findings will stimulate further research into Mn$_4$Al$_{11}$ and motivate advanced experimental investigations, such as angle-resolved photoemission spectroscopy and neutron scattering, to uncover more about its intriguing properties.

\begin{acknowledgments}
This work was supported by National Natural Science Foundation of China (Grant No.12274033, 12488201), and National Key Research and Development Program of China (2022YFA1403901, 2024YFA1408700). 
\end{acknowledgments}



\bibliography{Ref}

\end{document}